\documentclass[aps,pre,psfig,twocolumn,showpacs,address,nobibnotes,
reprint]{revtex4-1}
\usepackage{color}
\usepackage{graphics}
\usepackage{epsfig}
\usepackage[normalem]{ulem}
\usepackage{cancel}
\usepackage{float}
\usepackage{amsmath}
\usepackage{amsfonts}

\usepackage[outercaption]{sidecap}

\begin{document}
\title{Ionization and transport in partially ionized multicomponent plasmas: Application to atmospheres of hot Jupiters}
\author{Sandeep Kumar$^1$}
\email{sandeep.kumar@uni-rostock.de}
\author{Anna Julia Poser$^1$}
\author{Manuel Sch\"{o}ttler$^1$}
\author{Uwe Kleinschmidt$^1$}
\author{Wieland Dietrich$^2$}
\author{Johannes Wicht$^2$}
\author{Martin French$^1$}
\author{Ronald Redmer$^1$}
\affiliation{
$^1$Universit\"at Rostock, Institut f\"ur Physik, D-18051 Rostock, Germany\\ 
$^2$Max-Planck-Institut f\"ur Sonnensystemforschung, D-37077 G\"ottingen, Germany}
\date{\today}
\begin{abstract} 
\paragraph*{} 
We study ionization and transport processes in partially ionized multicomponent plasmas. The plasma composition is calculated via a system of coupled mass action laws. The electronic transport properties are determined by the electron-ion and electron-neutral transport cross sections. The influence of electron-electron scattering is considered via a correction factor to the electron-ion contribution. Based on this data, the electrical and thermal conductivity as well as the Lorenz number are calculated. For the thermal conductivity, we consider also the contributions of the translational motion of neutral particles and of the dissociation, ionization, and recombination reactions. We apply our approach to a partially ionized plasma composed of hydrogen, helium, and a small fraction of metals (Li, Na, Ca, Fe, K, Rb, Cs) as typical for hot Jupiter atmospheres. We present results for the plasma composition and the transport properties as function of density and temperature and then along typical $P$-$T$ profiles for the outer part of the hot Jupiter HD~209458b. The electrical conductivity profile allows revising the Ohmic heating power related to the fierce winds in the planet's atmosphere. We show that the higher temperatures suggested by recent interior models could boost the conductivity and thus the Ohmic heating power to values large enough to explain the observed inflation of HD~209458b. 
\end{abstract} 

\maketitle 
  
\section{Introduction}
\label{intro}

Partially ionized multicomponent plasmas are composed of molecules, atoms, and ions of various species as well as of free electrons. The plasma parameters density and temperature determine their ionization degree and, thus, also their equation of state and transport properties like electrical and thermal conductivity~\cite{Trapo_1983, GuRa_1984, Kraeft1986, Fortov1990, Redmer_1997, Likalter_1997, RESOS_2016, Zaghloul_2020}. Profound knowledge of the thermophysical properties of such plasmas is important for applications in astrophysics, atmospheric science, and plasma technology. For instance, Earth's ionosphere~\cite{Kallenrode, JMW_2009} and the atmospheres of hot Jupiters~\cite{Batygin_2010, Komacek_2017} can be treated as low-density partially ionized multicomponent plasmas. Another example is the formation of stars out of initially cold and dilute clouds which consist mostly of molecular hydrogen, helium, and a small fraction of heavier elements and collapse due to gravitational instability~\cite{Star-form, Inutsuka_2012}. The evolution to a protostar, which is much hotter and denser, runs through the plasma regime, where dissociation and ionization processes determine the heating and contraction dynamics essentially. The quenching gas in high-power circuit breakers~\cite{Franck_2006} or arc plasmas~\cite{Guo_2017} are examples of important technical applications of multicomponent partially ionized plasmas (PIP). 
% \RR{More examples? Add further citations!}

The transport properties of partially ionized plasmas are determined by the ionization degree and the charge state distribution of its constituents. This defines the number of free electrons and the strength of the collisional interactions between the plasma species and determines their mobility. At low temperatures, the ionization degree is very low and the transport properties are dominated by neutral particles (atoms, molecules), while charged particles become more and more important with increasing temperature due to thermal ionization. Such thermal ionization conditions are typical for the outer atmospheres of planets in close proximity to their star, like hot Jupiters and hot mini-Neptunes~\cite{Pu_2017}. The electrical conductivity, in particular due to the ionization of alkali metals, can rise to values where magnetic effects become important for the evolution and dynamics of the planetary interior. 

Hot Jupiters orbit their parent stars in close proximity and are locked in synchronous rotation, which means that they always face the same side to the star. Several physical mechanisms are discussed to explain why the radii of hot Jupiters are significantly larger than expected~\cite{Komacek_2017,Sarkis_2020,Burrows_2007}. One possibility is Ohmic dissipation that directly scales with the electrical conductivity.

The differential stellar irradiation drives fierce winds in the outer atmosphere that tend to equilibrate the difference in dayside and nightside temperature. Interaction of the winds with a planetary magnetic field induces electric currents that can flow deeper into the planet. When efficient enough, the related Ohmic heating transports a sufficient fraction of the stellar irradiation received by the planet to deeper interiors where it could explain the inflation.

Accurate data on the composition and the transport coefficients along realistic pressure-temperature ($P$-$T$) profiles of hot Jupiters are also critical input in corresponding magnetohydrodynamics simulations~\cite{Rogers_2014}. Using the corresponding plasma composition, i.e., the molar fractions of the various species, and the absorption coefficient of the plasma, the opacity of the planet's atmosphere can be calculated, which, in turn, determines the $P$-$T$ profile~\cite{Freedman_2008}.

In this paper, we calculate the ionization degree, the electrical and thermal conductivity, and the Lorenz number for a PIP as function of temperature and mass density. Mass-action laws (MALs) are used to calculate the composition of the PIP~\cite{Redmer_1988, Redmer_1999, Kuhlbrodt_2000, Kuhlbrodt_2005, Schottler_2013}. We assume that the plasma is in thermal and chemical equilibrium so that Saha-like equations for each dissociation and ionization reaction can be derived, from which the partial densities of all species are calculated, i.e., the plasma composition. Furthermore, the electron-ion and electron-neutral transport cross sections have to be determined~\cite{Martin_2017}. The effect of electron-electron scattering is considered by introducing a correction factor to the electron-ion contribution according to the Spitzer theory~\cite{Spitzer_1953}. Note that the influence of the electron-electron interaction on the transport coefficients is currently of interest also for dense, non-ideal plasmas~\cite{Reinholz_2015, Desjarlais_2017}. The contribution of the translational motion of neutrals and of the heat of dissociation, ionization, and recombination reactions to the thermal conductivity of PIP was also studied. For a benchmark, we have compared the thermal conductivity of hydrogen plasma obtained from our model to the experimental arc-discharge results of Behringer and van Cung~\cite{behringer_1980}. In a next step, we study the general trends of the ionization degree and of the transport coefficients with respect to the plasma density and temperature. Finally, we calculate the ionization degree as well as the electrical and thermal conductivity along typical $P$-$T$ profiles through the atmosphere of the inflated hot Jupiter HD~209458b. These results are then used to assess the Ohmic heating in the planet's atmosphere and to infer whether this effect is efficient enough to explain the inflation. Batygin and Stevenson~\cite{Batygin_2010} (referred to as B\&S10 from now on) have used simplified expressions for the calculation of the plasma composition (ionization scaled with the density scale height) and the electrical conductivity (weakly ionized gas) and concluded that Ohmic heating is indeed sufficient to explain the inflation of this hot Jupiter. We use our refined conductivity values to calculate updated estimates for the Ohmic heating in HD~209458b. 
 
Our paper is organized as follows. In Section~\ref{basic} we outline the theoretical basics for the calculation of the equation of state (EOS) and the composition of the PIP. Section~\ref{tran_coe} provides the basic formulas used for the calculation of the electronic transport coefficients in the PIP. In Section~\ref{result_transport}, we report the results for the ionization degree and the electronic transport coefficients in dependence of the plasma temperature and mass density. Section~\ref{tr_react_heat} gives details of the calculation of the translational motion of neutral particles and of the contribution of the heat of dissociation, ionization, and recombination reactions to the thermal conductivity. In Section~\ref{jupiter}, results for the ionization degree and the transport coefficients along typical $P$-$T$ profiles through the atmosphere of the hot Jupiter HD~209458b are presented. Conclusions are given in Section~\ref{concl}.

%%%%%%%%%%%%%%%%%%%%%%%%%%%%%%%%%%%%%%%%%%%%%%%%%
\section{Equation of state (EOS) and composition}
\label{basic}
%\subsection{}

We consider an ideal-gas-like model % (IGM) RR: never used 
for the partially ionized plasma and calculate its chemical composition using a canonical partition function $Z(\left\{N_i\right\}, V, T)$, which depends on the number of particles $N_i$ of species $i$ as well as on the volume $V$ and temperature $T$ of the plasma. We assume the constituent elements H, He, Li, K, Na, Rb, Ca, Fe, and Cs to be the relevant drivers of ionization in hot Jupiter atmospheric plasmas. The abundance of these constituents is given in Table~\ref{abund_para}, which is adopted from Refs.~\cite{Goswami, Lodders_2003}.  

\begin{table}[!hbt]
\caption{Abundance of constituents considered in this work: molar and mass fraction according to Refs.~\cite{Goswami, Lodders_2003}.}
\begin{center}
\begin{tabular}{ccc}
\hline\hline
element & molar fraction [\%] & mass fraction [\%] \\ %[1.5ex] 
\hline
 H  & $92.23$ & 74.84 \\ 
 He & $7.76$  & 25.02 \\
 Li & $1.75\times 10^{-7}$  & $9.82\times 10^{-7}$ \\
 Na & $1.79\times 10^{-4}$  & $3.3\times 10^{-3}$ \\
 K  & $1.18 \times 10^{-5}$ & $3.74\times 10^{-4}$ \\
 Ca & $1.88\times 10^{-4}$  & $6.1\times 10^{-3}$ \\
 Fe & $2.7\times 10^{-3}$   & $0.119$ \\
 Rb & $2.21\times 10^{-8}$  & $1.52\times 10^{-6}$ \\
 Cs & $1.16\times 10^{-9}$  & $1.23\times 10^{-7}$ \\
\hline\hline
\end{tabular}
\end{center}
\label{abund_para}
\end{table} 

In a mixture of $c$ non-interacting chemical species, the partition
function $Z(\left\{N_i\right\}, V, T)$ can be written as a product:
  \begin{equation}
     {Z(\left\{N_i\right\}, V, T) = \displaystyle\prod_{i=1}^{c} z_i(N_i, V, T) }\, ,
     %\nonumber
    \end{equation}
with   
    \begin{equation}
     {z_i(N_i, V, T) = z_i^{trans}(N_i, V, T) (z_i^{int}(T))^{N_i}}\, ,
     %\nonumber
    \end{equation}
where $z_i^{trans}(N_i, V, T)$ is the translational partition function of species $i$ and $z_i^{int}(T)$ is its one-particle internal partition function (IPF). The translational partition function is given by:
\begin{equation}
     {z_i^{trans}(N_i, V, T) = \frac{V^{N_i}}{N_i !\lambda_{th, i}^{3N_i}}   } \, ,
     %\nonumber
    \end{equation}
in which $\lambda_{th, i} = h/\sqrt{2\pi m_i k_B T}$ is the thermal wavelength with the Planck constant $h$, the mass $m_i$ of species $i$, and the Boltzmann constant $k_B$. The internal partition function modes are considered to be independent from each other, which gives the following formula~\cite{Patharia_2011_141}:
\begin{equation}
     z_i^{int} = z^{nuc}_i z^{el}_i z^{vib}_i z^{rot}_i \, ,
     %\nonumber
    \end{equation}
where $z^{nuc}_i$, $z^{el}_i$, $z^{vib}_i$, and $z^{rot}_i$ are the nuclear, electronic, vibrational, and rotational partition functions of the species $i$, respectively. 

The nuclear IPF is considered as follows: 
\begin{equation}
     {z^{nuc}_i = 2I^{ns}_i+1} \, ,
    % \nonumber
    \end{equation}
which depends on the spin quantum number $I^{ns}_i$ of the nucleus. The electronic partition function is approximated as follows:
\begin{equation}
     {z^{el}_i = (2J+1)\exp\left(-E_i^0/k_B T\right)} \, .
    % \nonumber
    \end{equation}
Here, $E_i^0$ is the energy and $J$ the electronic angular momentum quantum number of the atom/ion/molecule in the ground state. We do not consider excited states in this study because their population is small for the plasma parameters considered here so that their effect on ionization and transport is negligible. Note that each excited state introduces a new species for which all related atomic/ionic/molecular parameters need to be known for the calculation of the plasma composition and the transport cross sections which would unnecessarily complicate the PIP model as long as their effect is small. For the calculation of the vibrational and rotational partition function of the $\textrm{H}_2$ molecule we use the high-temperature approximation,
%. The expressions are the following:
    \begin{eqnarray}
     z^{vib}_{\textrm{H}_2} &=& \frac{1}{[1-\exp(-\theta_v /T)]} \,,
     \label{vib_pf} \\
%    \end{equation}
%    \begin{equation}
     z^{rot}_{\textrm{H}_2} &=& \frac{T}{2\theta_r} \,,
     \end{eqnarray}
where $\theta_v = h\nu/k_B$ and $\theta_r = h^2/8\pi^2 Ik_B$ are the vibrational and rotational temperature, respectively. The latter depends on the moment of inertia $I = \mu_{\textrm{HH}} r^2$ of the $\textrm{H}_2$ molecule; $\mu_{\textrm{HH}}$ = $m_\textrm{H}$/2 is its reduced mass and $r$ its bond length. 

The plasma considered here is in thermal and chemical equilibrium, so that the particle densities follow from MALs as follows~\cite{Schottler_2013}:
  \begin{equation}
     {\displaystyle\prod_{i} n_i^{\nu_{i,a}} = \displaystyle\prod_{i} \frac{(z_i^{int})^{\nu_{i,a}}}{(\lambda_{th, i}^{3})^{\nu_{i,a}}} \equiv K_a(T)} \,.
     \label{Saha_eq}
  \end{equation}
In this expression, $n_i=N_i/V$ are number densities of species $i$, $K_a(T)$ is the reaction constant, and $\nu_{i,a}$ are the stoichiometric coefficients of the reaction $a$. The $\nu_{i,a}$ for the reaction products and reactants are chosen to be positive or negative, respectively. The MALs and particle conservation equations of the PIP were solved numerically to calculate the number density of each species for a given temperature and mass density. We allowed the constituents to be doubly ionized at maximum. This sets a maximum temperature of about $30~000$~K for our applications, which corresponds to about 10~\% of the lowest third ionization energy (30.651~eV for Fe~\cite{NIST_ASD}) of all constituents considered. The MALs for dissociation and ionization read as follows:
  \begin{eqnarray}
    \frac{n_\textrm{H}^2}{n_{\textrm{H}_2}} &=& K_\textrm{H} \, ,
    \label{diss_eq} \\
%  \end{equation}
%  \begin{equation}
    \frac{n_{ion,i}^{+} n_e}{n_{atom,i}} &=& K_{ion,i}^{+} \, ,
    \label{single_iz_eq} \\
%  \end{equation}
%  \begin{equation}
    \frac{n_{ion,i}^{2+} n_e}{n_{ion,i}^{+}} &=& K_{ion,i}^{2+} \, ,
    \label{doubly_iz_eq}
  \end{eqnarray}
in which $n_{ion}^{+}$ and $n_{ion}^{2+}$ denote the number densities of a singly or doubly charged ion, respectively. The charge neutrality condition in the PIP leads to the following equation:  
    \begin{equation}
     {n_e = \sum_{i} n_{ion,i}^{+} + \sum_{i} 2n_{ion,i}^{2+}} \, ,
     \label{chg_neu_eq}
    \end{equation}
where $n_e$ represents the free electron number density in the PIP. Mass conservation in the plasma provides the following relation:
   \begin{equation}
     {\rho = \sum_{i=1}^{c} m_i n_i} \,  ,
    \label{mass_cons_eq}
    \end{equation}
where $\rho$ is mass density of the plasma. The relative abundance $\chi_r$ of each constituent with respect to the $\textrm{H}$ abundance has been set as follows:     
    \begin{equation}
     {\chi_{r,i} = \frac{n_{atom,i} + n_{ion,i}^{+} + n_{ion,i}^{2+}}{2n_{\textrm{H}_2} + n_\textrm{H} + n_{\textrm{H}^{+}}}} \, .
    \label{abundz_eq}
    \end{equation}
    
Most of the parameters like ground state energies $E_i^0$, ionization energies, total angular momentum quantum numbers $J$, and atomic weights $m_i$ of the species are taken from the NIST database~\cite{NIST_ASD}. The nuclear partition function and ground state energy of the $\textrm{H}_2$ molecule are taken as $z_{\textrm{H}_2}^{nuc}$ = $4$ and $E_i^0 = -31.738$~eV~\cite{Schottler_2013}, respectively. The ground state energy of the H$_2$ molecule already includes the vibrational ground-state energy. Therefore, the vibrational partition function, Eq.~(\ref{vib_pf}), includes only excited states. We have taken $\theta_v = 6321.3$~K and $\theta_r = 88.16$~K for the vibrational and rotational temperature of the $\textrm{H}_2$ molecule~\cite{Bosnjakovic_1998}, respectively. The number densities $n_i$ of each species (molecules, atoms, ions) for a given plasma temperature and mass density were calculated by solving the coupled Eqs.~(\ref{diss_eq}), (\ref{single_iz_eq}), (\ref{doubly_iz_eq}), (\ref{chg_neu_eq}), (\ref{mass_cons_eq}), and (\ref{abundz_eq}) %in a C++ progam package 
using the Newton-Raphson method. The resulting ionization degree $\alpha$ of the plasma is defined as 
    \begin{equation}
     {\alpha = \frac{n_e}{n_{total}}} \, ,
    \end{equation}
with $n_{total} = n_{atoms} + 2n_{\textrm{H}_2}+ n_e$ and the density of all atoms $n_{atoms} = \sum_{i} n_{i, atom}$.

\begin{figure}[!hbt] 
\centering   
\includegraphics[height = 8.0cm,width = 8.6cm]{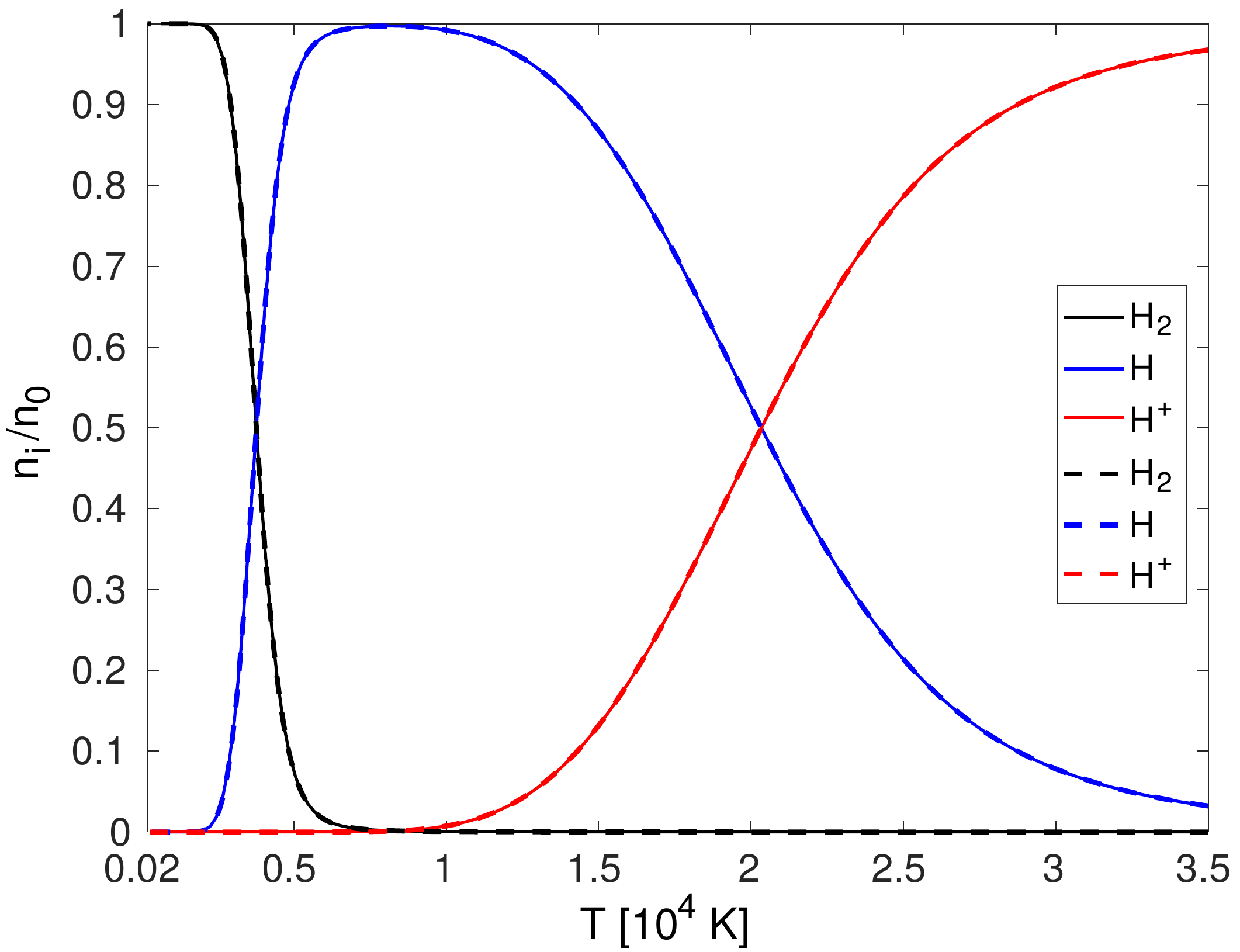}
\caption{Composition of hydrogen plasma as function of temperature for a mass density of $\rho=10^{-5}$~g/$\text{cm}^3$.
Solid lines: analytical results via Eqs.~(\ref{bench_eq_H}), (\ref{bench_eq_H2}), and (\ref{bench_eq_Hp}); Dashed lines: numerical results. The normalization $n_0 = n_{\textrm{H}_2} + n_\textrm{H} + n_{\textrm{H}^{+}}$ refers here to total number density of hydrogenic species in the plasma.}
\label{benchmark_den}	                           
\end{figure}

The numerical calculations were benchmarked against the analytical solution of Eqs.~(\ref{bench_eq_H}), (\ref{bench_eq_H2}), and (\ref{bench_eq_Hp}) for a pure hydrogen plasma composed of $\textrm{H}_2$, $\textrm{H}$, $\textrm{H}^{+}$ and electrons:
\begin{equation}
     {\frac{m_{\textrm{H}_2}}{K_{\textrm{H}_2}}n_\textrm{H}^2 + m_\textrm{H}n_\textrm{H} + m_{\textrm{H}^{+}}\sqrt{n_\textrm{H} K_{\textrm{H}}^{+}} -\rho = 0} \,.
     \label{bench_eq_H}
    \end{equation}
In the analytical model, the density of H atoms $n_\textrm{H}$ is obtained from the solution of Eq.~(\ref{bench_eq_H}). Furthermore, using $n_\textrm{H}$, we can calculate $n_{\textrm{H}_2}$ and $n_{\textrm{H}^+}$ via the following equations:
    \begin{eqnarray}
     n_{\textrm{H}_2} &=& \frac{n_\textrm{H}^2}{K_\textrm{H}} \,,
     \label{bench_eq_H2} \\
%    \end{equation}    
%    \begin{equation}
     n_{\textrm{H}^{+}} &=& \sqrt{n_\textrm{H} K_{\textrm{H}}^{+}} \,.
     \label{bench_eq_Hp}
    \end{eqnarray}
For benchmarking, the mass density $\rho$ of the plasma was kept constant at $10^{-5}$ g/$\text{cm}^3$ and the composition was calculated as function of the temperature; see Fig.~\ref{benchmark_den}. The analytical and numerical results are virtually identical. At low temperatures, hydrogen is a molecular gas; the molecules dissociate into atoms with increasing temperature. At even higher temperature, the ionization processes lead to a hydrogen plasma. For further validation, we have compared our results for the ionization degree with those of Schlanges {\it et al.}~\cite{Schlanges_1995} and found good agreement.

%%%%%%%%%%%%%%%%%%%%%%%%%%%%%%%%%%%%%%%%%%%%%%%%%%%%%%%%  

%%%%%%%%%%%%%%%%%%%%%%%%%%%%%%%%%%%%%%%%%%%
\section{Electronic Transport Coefficients}
\label{tran_coe}

The electronic contribution to the electrical conductivity $\sigma_e$, the thermal conductivity $\lambda_e$, and the Lorenz number $L$ are defined as follows~\cite{Lee_More_1984, Martin_2017}:
\begin{eqnarray}
 \sigma_e &=& e^2 K_0 \,,\\ 
 \lambda_e &=& \frac{1}{T}\Bigg(K_2 - \frac{K_1^2}{K_0}\Bigg) \,,\\ 
 L &=& \bigg(\frac{e}{k_B}\bigg)^2 \frac{\lambda_e}{\sigma_e T} \, ,
\end{eqnarray}
where $e$ is the elementary charge and $K_n$ are Onsager coefficients ($n$ = 0, 1, 2) that are composed of individual specific Onsager coefficients $K_{n,es}$ via:
\begin{equation}
{K_n^{-1} = \sum_{s}^{} K_{n,es}^{-1}} \, .
\label{convenient_form}
\end{equation}

The expressions for $K_{n,eN}$ and $K_{n,eI}$ are taken from French and Redmer~\cite{Martin_2017} and describe the contribution of electron scattering from neutral (index $N$) and ionic (index $I$) species, respectively. We have considered electron-neutral scattering only for $\textrm{H}$, $\textrm{H}_2$, and $\textrm{He}$ atoms/molecules because of the very small overall abundance of the heavier elements. The analytical expression of the specific Onsager coefficients for electron-neutral scattering for Eq.~(\ref{convenient_form}) is: 
\begin{equation}
 K_{n,eN} = \frac{2^{11/2}\pi^{1/2}(n+3)!~\varepsilon_0^2(k_BT)^{n + 3/2} n_e}{3Z_N^2e^4 m_e^{1/2}n_N \ln A_N(x_N)} \, .
 \label{anly_en}
\end{equation} 
The specific Onsager coefficients for electron-ion scattering for Eq.~(\ref{convenient_form}) read:
\begin{equation}
 K_{n,eI} = \frac{2^{11/2}\pi^{1/2}(n+3)!~\varepsilon_0^2(k_BT)^{n + 3/2} n_e}{3Z_I^2e^4 m_e^{1/2}n_I \ln \Lambda(B_n)} \, .
 \label{anly_ei}
\end{equation} 
The logarithmic functions $\ln A_N(x_N)$ and $\ln \Lambda(B_n)$ are defined in Ref.~\cite{Martin_2017}. Electron-electron scattering is accounted for by correction factors according to Spitzer and H\"{a}rm~\cite{Spitzer_1953} in the Onsager coefficients for electron-ion scattering $K_{n,eI}$. The respective formula and parameters are taken from French and Redmer~\cite{Martin_2017}. Recently, the effect of electron-electron scattering on the electrical and thermal conductivity of dense plasmas in the warm dense matter regime has been studied by Reinholz \textit{et al.}~\cite{Reinholz_2015} using linear response theory and by Dejarlais \textit{et al.}~\cite{Desjarlais_2017} using Kohn-Sham density functional theory. The expressions for the Onsager coefficients including electron-electron scattering are:
\begin{eqnarray}
 K_{0,eI+ee} &=& \frac{f_e}{f_I}K_{0,eI} \,,\\
%\end{equation}
%\begin{equation}
 K_{1,eI+ee} &=& \frac{a_ef_e}{a_If_I}K_{1,eI} \nonumber\\ 
  & & + \frac{5}{2}k_B T \Bigg( \frac{f_e}{f_I} - \frac{a_ef_e}{a_If_I} \Bigg) K_{0,eI} \,,\\
%\end{equation}
%\begin{equation}
 K_{2,eI+ee} &=& \frac{L_ef_e}{L_If_I}\Bigg(K_{2,eI} - \frac{K_{1,eI}^2}{K_{0,eI}}\Bigg) 
  + \frac{K_{1,eI+ee}^2}{K_{0,eI+ee}} \,,
\end{eqnarray}
where the factors $f_I$, $f_e$, $a_I$, $a_e$, $L_I$, and $L_e$ are defined in Ref.~\cite{Martin_2017}.

%%%%%%%%%%%%%%%%%%%%%%%%%%%%%%%%%%%%%%%%%%%%%%%%%
\section{Results for electronic transport in PIP}
\label{result_transport}

\begin{figure}[!hbt] 
\centering  
\includegraphics[height = 8.0cm,width = 8.5cm]{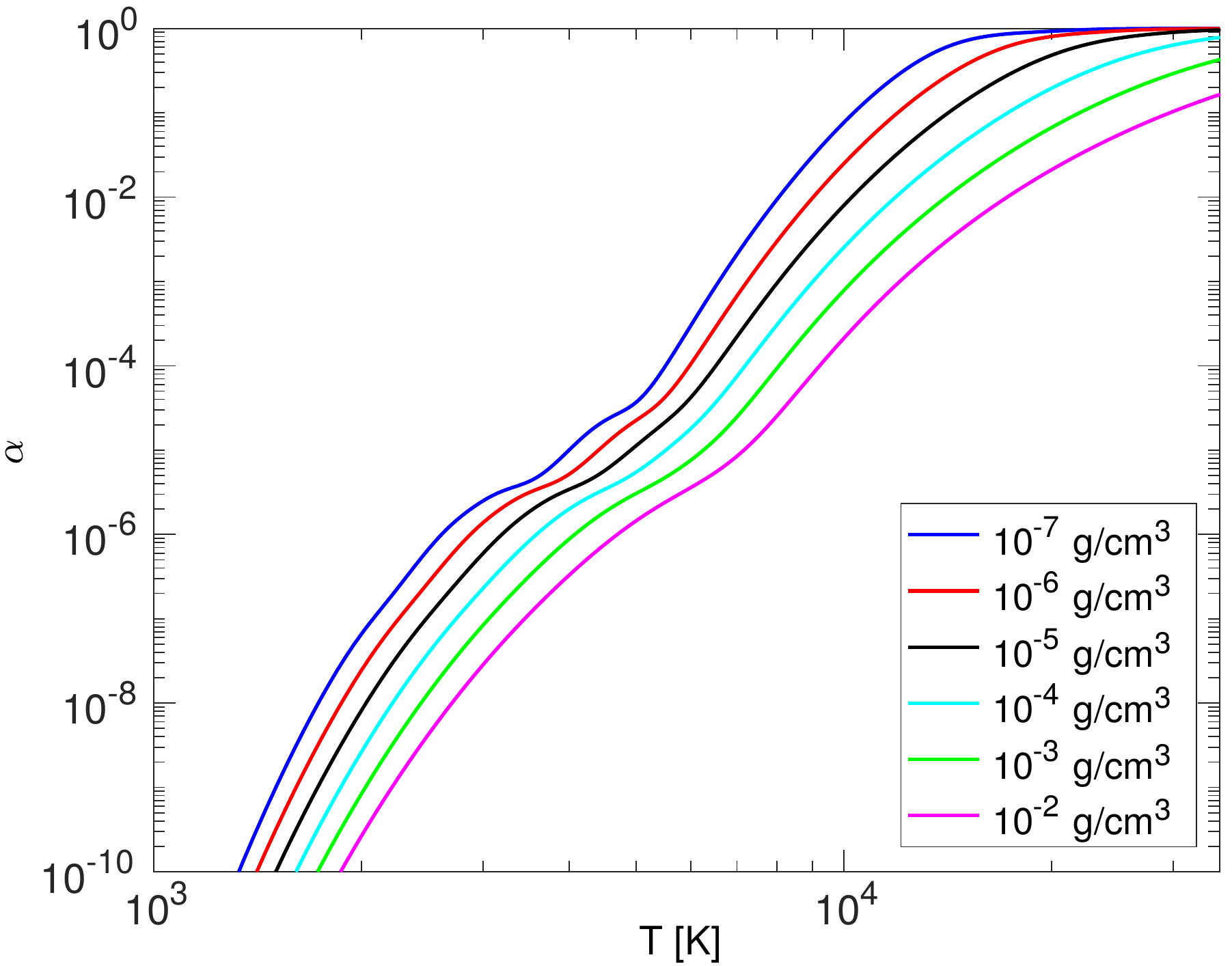}
\caption{Ionization degree of the PIP as function of temperature for different mass densities.}
\label{ID_vary_T}	                       
\end{figure}

The plasma composition, i.e., the partial number densities $n_i$ of each species obtained from solving the coupled Eqs.~(\ref{diss_eq}) -- %, (\ref{single_iz_eq}), (\ref{doubly_iz_eq}), (\ref{chg_neu_eq}), (\ref{mass_cons_eq}), and 
(\ref{abundz_eq}), is a necessary input for the calculation of the electronic transport coefficients. Therefore, we first show the behavior of the ionization degree as function of the temperature at different mass densities in Fig.~\ref{ID_vary_T}. The ionization degree $\alpha$ is increasing with the temperature due to thermal ionization of the constituents and decreasing with the mass density of the plasma.

%%%%%%%%%%%%%%%%%%%%%%%%%%%%%%%%%%%%%%%%%%%%%%%%%%%%%%%%%%%%%%%%%%%%

\begin{figure}[!hbt] 
\centering   
\includegraphics[height = 8.0cm,width = 8.5cm]{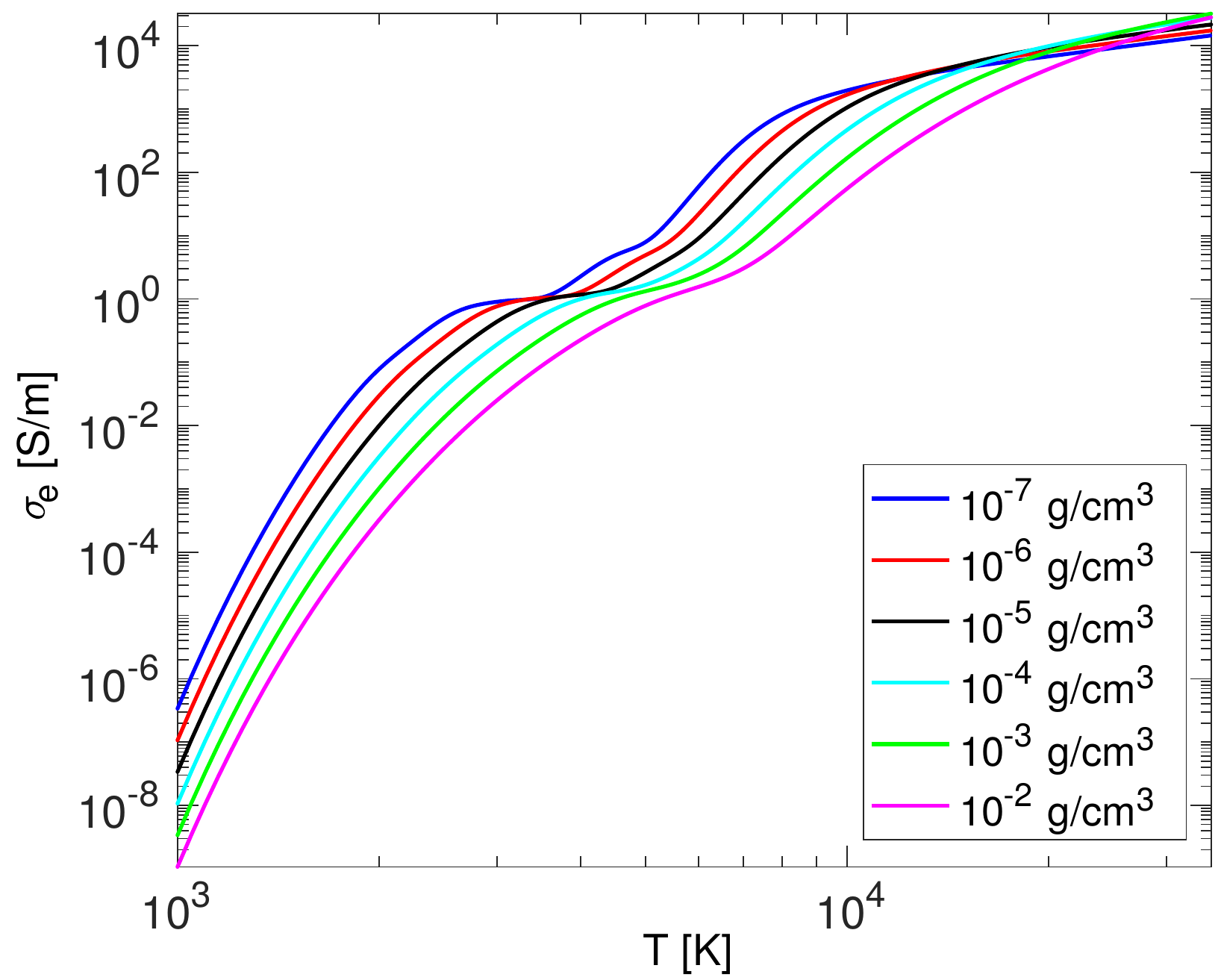}
\caption{Electrical conductivity of the PIP as function of temperature for different mass densities.}
\label{EC_vary_T}	                           
\end{figure}

\begin{figure}[!hbt] 
\centering   
\includegraphics[height = 8.0cm,width = 8.5cm]{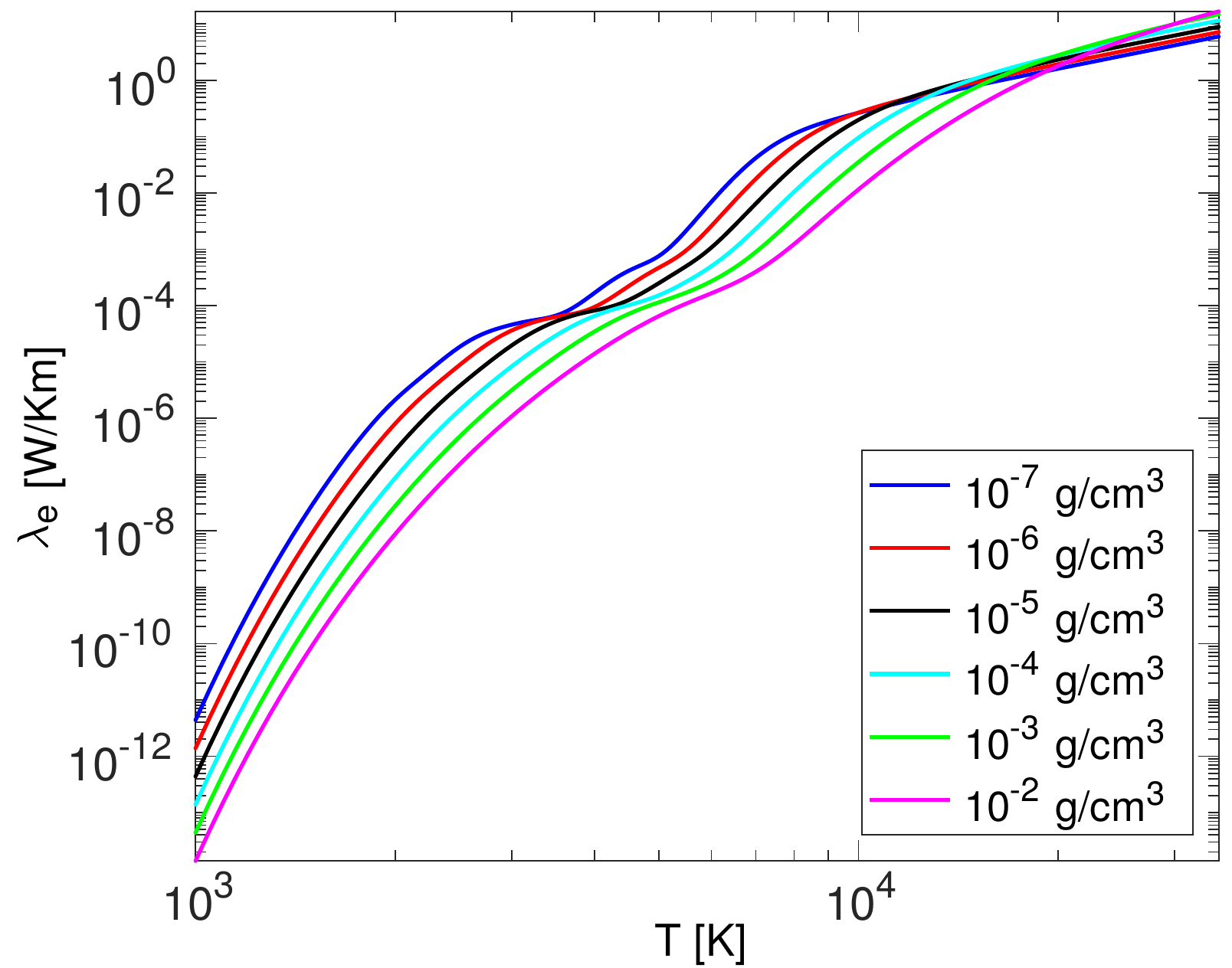}
\caption{Thermal conductivity of the PIP as function of temperature for different mass densities.}
\label{TC_vary_T}	                           
\end{figure}

\begin{figure}[!hbt] 
\centering   
\includegraphics[height = 8.0cm,width = 8.5cm]{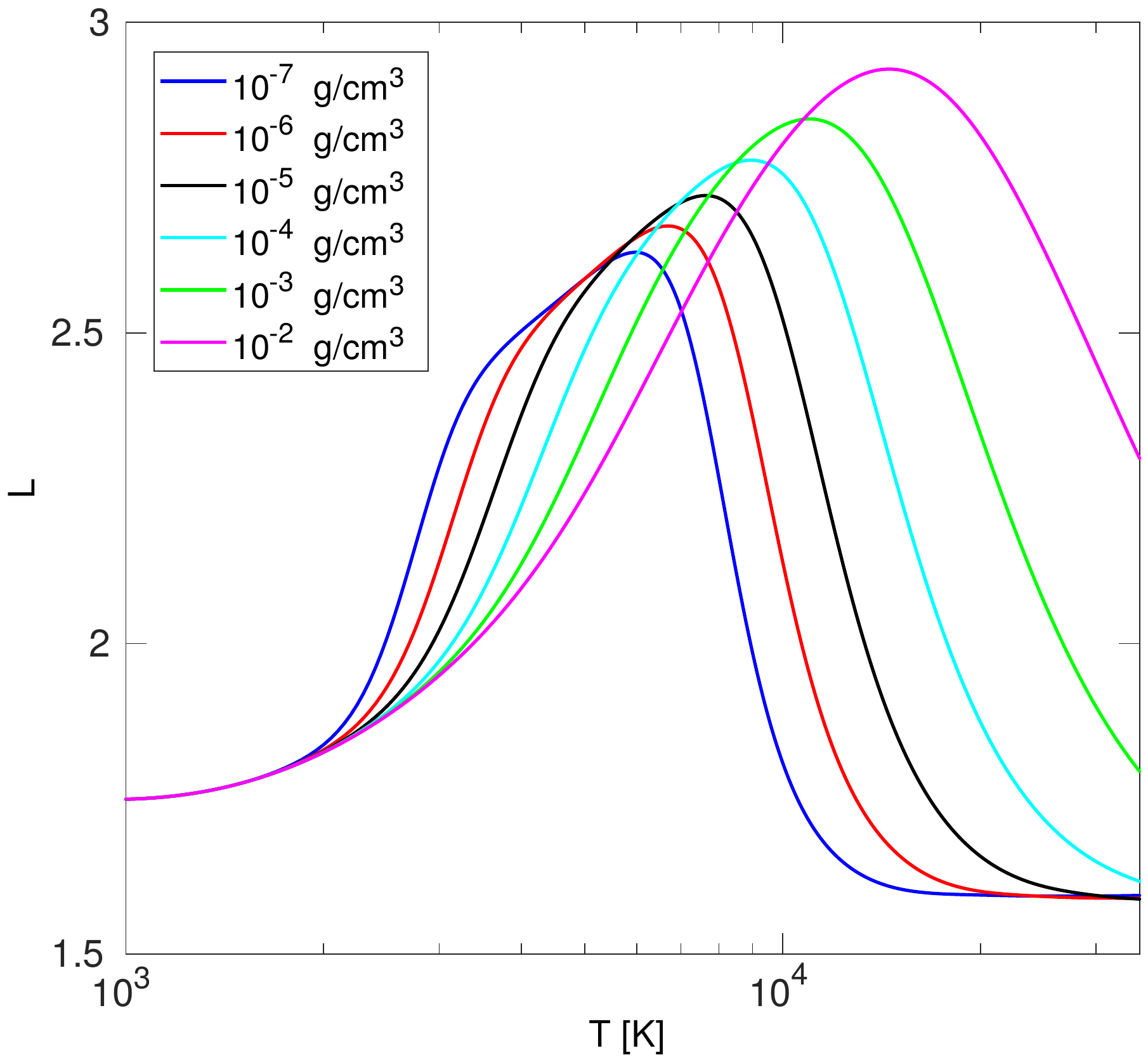}
\caption{Lorenz number of the PIP as function of temperature for different mass densities.}
\label{LN_vary_T}	                           
\end{figure}

The variation of $\sigma_e$ and $\lambda_e$ with the temperature at different mass densities is displayed in Figs.~\ref{EC_vary_T} and \ref{TC_vary_T}, respectively. The curves for $\sigma_e$ and $\lambda_e$ show systematic increase with temperature, caused by thermal ionization of the constituents in the order of their ionization energies which leads to an enhancement of the free electron density in the PIP. On the other hand, $\sigma_e$ and $\lambda_e$ are decreasing with mass density due to more frequent scattering processes with neutral species. At high temperatures (above 20~000 K), $\sigma_e$ and $\lambda_e$ are increasing with mass density, oppositely to their low-temperature characteristics. This reversal is emerging because the ionization degree is still increasing with temperature for the higher densities but it is already saturated for the lower densities. The quantities $\alpha$, $\sigma_e$, and $\lambda_e$ show plateau-like structures. When the plateau is reached, all metals (see Table~\ref{abund_para}) are ionized but H and He require still higher temperatures to contribute to the ionization degree significantly and thus to the electrical and thermal conductivity, which leads to the increase after the plateau. 
The Lorenz number shown in Fig.~\ref{LN_vary_T} first increases with the temperature and, after passing through a maximum, decreases for still higher temperatures. This behavior is shifted systematically towards higher temperatures with increasing density. The high- and low-temperature limiting values of $L$ are determined by the known Spitzer limit in the fully ionized plasma and electron-neutral cross sections in the weakly ionized gas, respectively. The occurrence of the pronounced maximum in $L$ is caused by different energetic weightings of the cross sections in the specific Onsager coefficients; see Eq.~(\ref{anly_en}) and (\ref{anly_ei}). It should be noted that the correction due to electron-electron scattering is only important when the majority of constituent elements are at least singly ionized.

%%%%%%%%%%%%%%%%%%%%%%%%%%%%%%%%%%%%%%%%%%%%%%%%%%%%%%%%%%%%%%%%%%%%        

%%%%%%%%%%%%%%%%%%%%%%%%%%%%%%%%%%%%%%%%%%%%%%%%%%%%%%%%%%%%%%%%%%%%        

%%%%%%%%%%%%%%%%%%%%%%%%%%%%%%%%%%%%%%%%%%%%%%%%%%%%%%%%%%%%%%%%%%%%        

%%%%%%%%%%%%%%%%%%%%%%%%%%%%%%%%%%%%%%%%%%%%%%%%%%%%%%%%%%%%%%%%%%%
\section{Thermal conductivity from neutrals and chemical reactions}
\label{tr_react_heat}

At low temperatures, the ionization degree is small and, therefore, the neutral particles contribute significantly to the heat transport. In addition to their translational contribution $\lambda_{tr}$, the occurrence of dissociation and ionization reactions also enhances the thermal conductivity in the corresponding temperature region, described by a term $\lambda_r$. These contributions have to be added to the electronic heat conductivity $\lambda_e$ so that the total thermal conductivity $\lambda$ of the PIP is given by:
\begin{equation}\label{eq:lambda}
\lambda = \lambda_e + \lambda_{tr} + \lambda_r \,.
\end{equation} 
We have neglected the translational contribution of ions to the thermal conductivity because it is very small in comparison to that of the electrons $\lambda_e$~\cite{Imshennik_1962, Reinholz_1989}. For the neutrals, we have adopted the Chapman-Enskog model for the calculation of the translational heat transport. The first-order expression for $\lambda_{tr}$ for a single-component gas is given by~\cite{maitland_1981}:
\begin{equation}
\lambda_i^{tr} = \frac{25}{32}\bigg(\frac{k_B T \pi}{m_i}\bigg)^{1/2} \frac{C_{v,i}}{\Omega^{(2,2)}_{ii}(T)} \, , 
\end{equation}
where $\Omega_{ii}^{(2,2)}(T)$ is a collision integral and $C_{v,i}$= $3k_B/2$ is the heat capacity for atoms of species $i$ at constant volume. The collision integral depends upon the energy-dependent transport cross section. We have simplified the collisional integral by assuming the atoms or molecules to be rigid spheres of diameter $d_{ii}$, so that $\Omega^{(2,2)}_{ii}(T)$ becomes temperature independent and is reduced to $\pi d_{ii}^2$. The simplified formula of $\lambda_i^{tr}$ is then given by
\begin{equation}
\lambda_i^{tr} = \frac{25}{32}\bigg(\frac{k_B T\pi}{m_i}\bigg)^{1/2} \frac{C_{v,i}}{\pi d_{ii}^2} \, .
\end{equation}
The vibrational heat capacity $C_{v,\textrm{H}_2}^{vib}$ of hydrogen molecules is calculated using the harmonic approximation~\cite{Patharia_2011_141}:
\begin{equation}
C_{v,\textrm{H}_2}^{vib} = k_B\left(\frac{\theta_v}{T}\right)^2\frac{\exp\left(\frac{\theta_v}{T}\right)}{\left[\exp\left(\frac{\theta_v}{T}\right) - 1\right]^2} \, .
\end{equation}
The rotational heat capacity $C_{v,\textrm{H}_2}^{rot}$ of hydrogen molecules is calculated by considering $T \gg \theta_r$ so that
\begin{equation}
C_{v,\textrm{H}_2}^{rot} = k_B \, .
\end{equation} 

For a multicomponent plasma as considered here we use a generalized formula for the calculation of the translational thermal conductivity of mixtures which reads~\cite{Mason_1958, brokaw_1958_mixture, Saksena_1967}:
\begin{equation}
\lambda_{tr} = \sum_{i} \frac{x_i \lambda^{tr}_i}{1 + \sum_{j \neq i} \frac{x_j}{x_i}\phi_{ij}} \, ,
\end{equation} 
where $x_i$ is the molar fraction of species $i$ and $\phi_{ij} = (2 \mu_{ij}/m_i)^2$ depends on the reduced mass $\mu_{ij}$ and mass of species $i$.

In the calculation of $\lambda_r$ we assume that the chemical reactions occur in different temperature regions, so that their contributions are additive, according to an expression given by Brokaw~\cite{Brokaw_1960, butler_1957}:
 \begin{equation}
\lambda_r =\sum\limits_a \frac{(\Delta H_a)^2}{RT^2} \frac{1}{A_a} \, ,
\end{equation}    
with
\begin{equation}
A_a = \sum_{k=1}^{\beta - 1} \sum_{l=k+1}^{\beta} \bigg(\frac{RT}{PD_{kl}}\bigg)x_k x_l \left[ \bigg(\frac{\nu_{k,a}}{x_k}\bigg) - \bigg(\frac{\nu_{l,a}}{x_l}\bigg) \right]^2 \, ,
\end{equation}
where $\Delta H_a$ is the heat of the reaction $a$, $\beta$ is the number of species involved in the reaction, $k$ represents the $k$th species, $R$ is the universal gas constant, $P=\sum_i n_ik_BT$ the ideal pressure, and $D_{kl}$ is the binary diffusion coefficient between components $k$ and $l$.  The heat of the reaction $\Delta H_a$ is calculated from the reaction constant by van't Hoff's equation~\cite{Hanley_1970, Kremer_2010}:
\begin{equation}
\frac{\Delta H_a}{RT^2} = \frac{d \ln K_a}{d T} \, .
\end{equation}
We use the following expression for the neutral-neutral and neutral-ion binary diffusion coefficients~\cite{hirschfelder1964}:
\begin{equation}
PD_{kl} = \frac{3}{16}\frac{\sqrt{2\pi k_B^3T^3/\mu_{kl}}}{\pi d_{kl}^2} \, .
\label{pd_eq}
\end{equation} 

For the electron-neutral and electron-ion diffusion coefficients, we have used the Darken relation and the adiabatic approximation~\cite{Hansen_1990}, which leads to:
\begin{equation}
D_{ke} = x_kD_e + x_e D_k \approx x_kD_e \, .
\end{equation}
This expression depends only on the self-diffusion coefficient $D_e$ of the electrons that can be related to their electrical conductivity using the Nernst-Einstein relation: 
\begin{equation}
PD_{ke} = \frac{x_k}{x_e}\bigg(\frac{k_B T}{e}\bigg)^2\sigma_e \, .
\end{equation}

%%%%%%%%%%%%%%%%%%%%%%%%%%%%%%%%%%%%%%%%%%%%%%%%%%%%
\begin{figure}[!hbt] 
\centering   
\includegraphics[height = 8.0cm,width = 8.5cm]{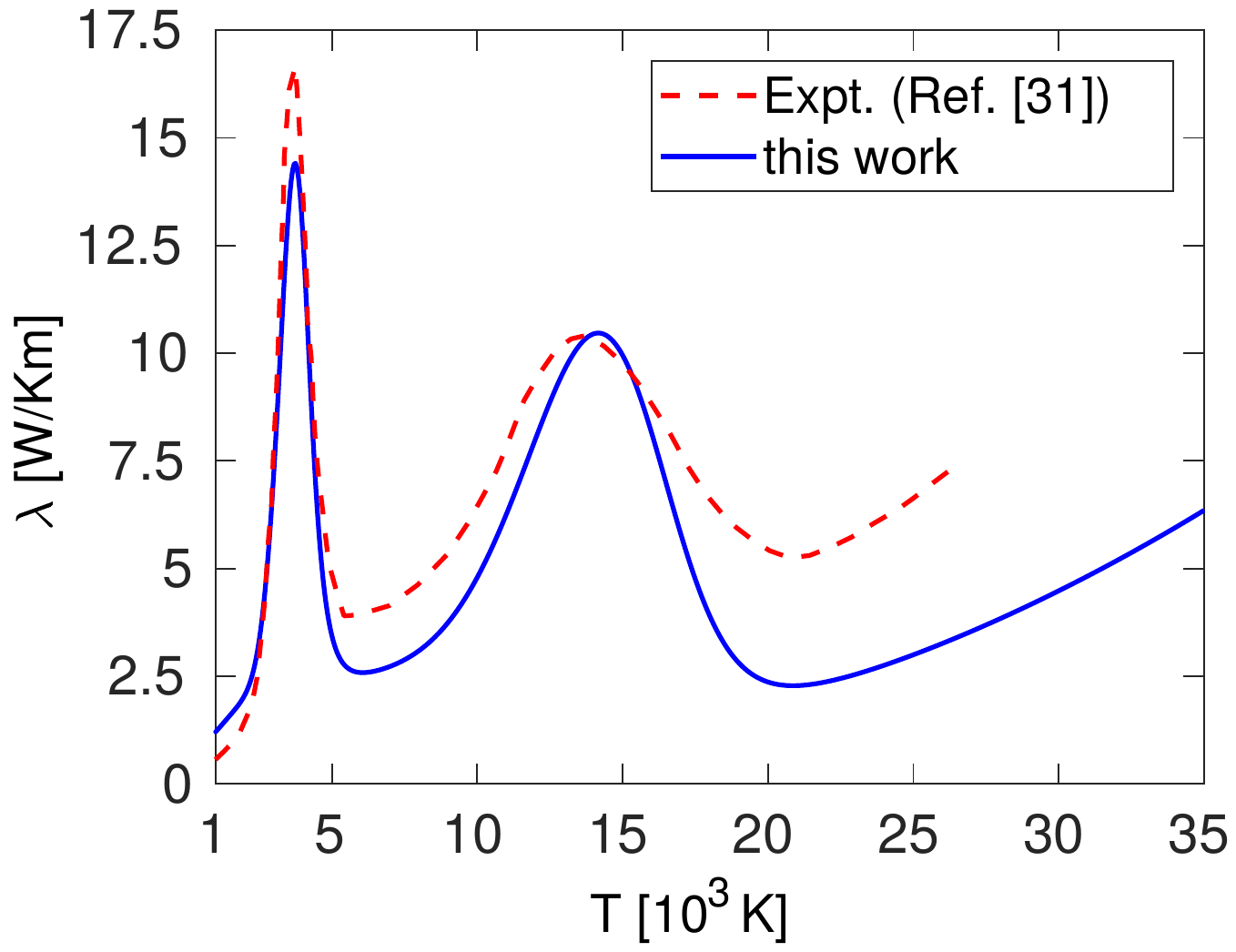}
\caption{Thermal conductivity of partially ionized hydrogen plasma as function of temperature at constant pressure of 1~bar. We compare our results with the arc discharge experiment of Behringer and van Cung~\cite{behringer_1980} which was evaluated using the local thermodynamic equilibrium assumption.} 
\label{exp_comp}	                           
\end{figure}
%%%%%%%%%%%%%%%%%%%%%%%%%%%%%%%%%%%%%%%%%%%%%%%%%%%%
\begin{figure}[!hbt] 
\centering   
\includegraphics[height = 8.0cm,width = 8.5cm]{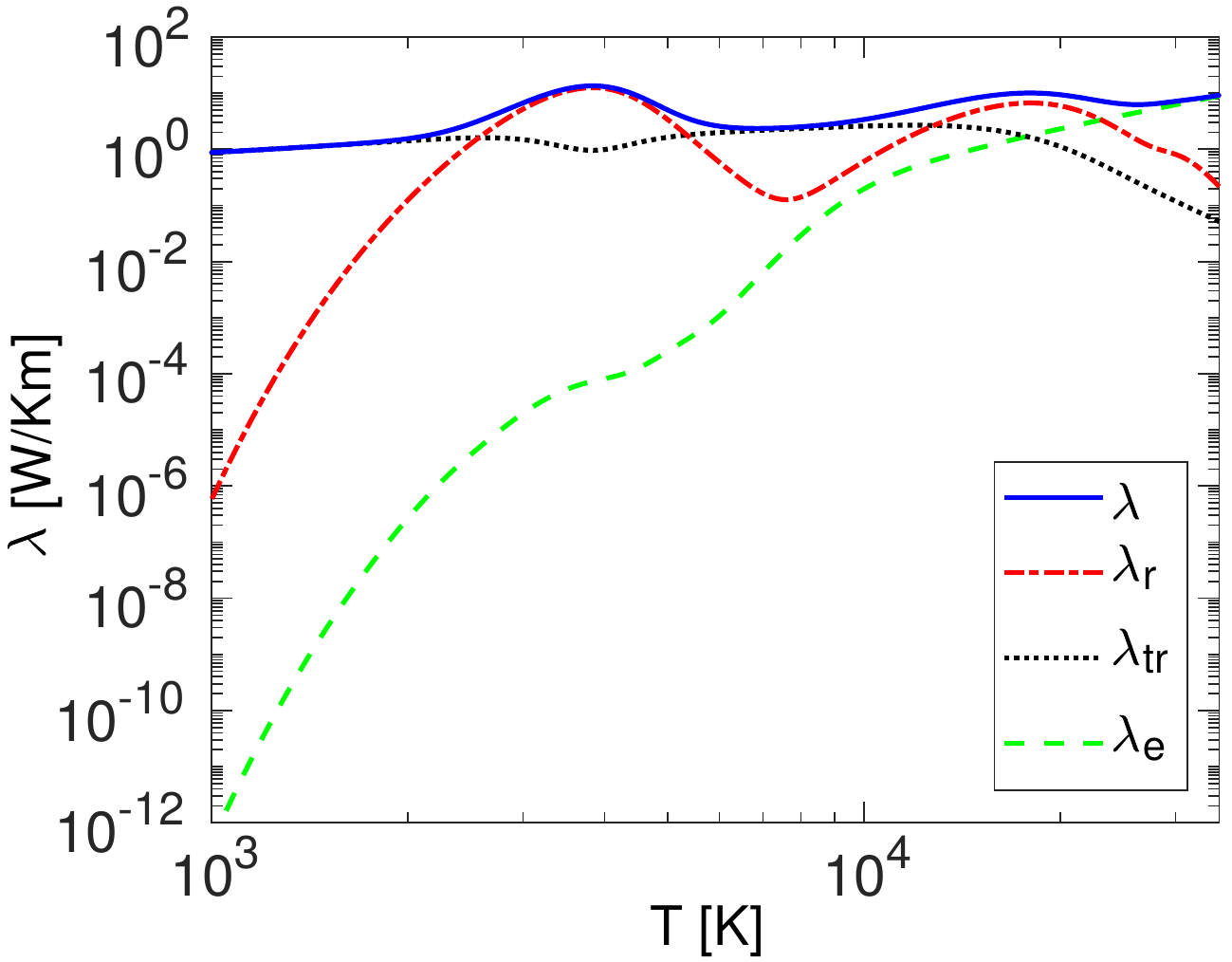}
\caption{Total thermal conductivity $\lambda$ according to Eq.~(\ref{eq:lambda}) as function of temperature at $10^{-5}$~g/cm$^3$. The contributions of the translational motion of neutrals $\lambda_{tr}$, of the heat of chemical reactions $\lambda_r$, and the electronic contribution are shown separately.}
\label{TC_all}	                           
\end{figure}
%%%%%%%%%%%%%%%%%%%%%%%%%%%%%%%%%%%%%%%%%%%%%%%%%%%%        

We have considered the $\lambda_{tr}$ and $\lambda_r$ contributions to the thermal conductivity only for species and reactions containing the elements H and He. The hard sphere diameters of $\textrm{H}_2$, H, and He are taken from Table~II in Vanderslice \textit{et al.}~\cite{vanderslice_1962}, specifically of $\textrm{H}$-$\textrm{H}_2$ collision data at 3500~K. We have parametrized the effective $\textrm{H}$-$\textrm{H}^{+}$ interaction diameter in our model by matching the height of the second peak in the thermal conductivity profile with that from hydrogen arc discharge experiments at $P=1$ bar~\cite{behringer_1980}; the comparison is shown in Fig.~\ref{exp_comp}. The $\textrm{He}$-$\textrm{He}^{+}$ and $\textrm{He}^{+}$-$\textrm{He}^{2+}$ interaction diameters have been calculated from Eq.~(\ref{pd_eq}) by using the diffusion coefficient value of Devoto and Li at 24~000~K~\cite{devoto_1968_He}. All hard sphere diameter values used for the calculation of the thermal conductivity are compiled in Table~\ref{dia_table}. 
\begin{table}[!hbt]
\caption{Square of the hard sphere diameters $d_{ij}$ for the interactions between the various species as used in the calculation of the thermal conductivity.}
\begin{center}
\begin{tabular}{ccc}
 \hline\hline 
 \rule{0pt}{2.5ex} collision & \quad $d_{ij}^2$~$[\AA^2]$ \\ %[1.5ex] 
 \hline
  $\textrm{H}_2$-$\textrm{H}_2$ & 2.634 \\ 
  H-$\textrm{H}_2$ & 2.634 \\
  H-H & 2.634 \\ 
  H-$\textrm{H}^{+}$ & 11.00 \\
  He-He & 2.634 \\
  He-$\textrm{He}^{+}$ & 13.978 \\
  $\textrm{He}^{+}$-$\textrm{He}^{2+}$ & 13.978 \\
 \hline\hline
\end{tabular}
\end{center}
\label{dia_table}
\end{table} 

The variation of $\lambda$, $\lambda_r$, $\lambda_{tr}$, and $\lambda_e$ with the temperature is displayed in Fig.~\ref{TC_all}, again for a constant density of $10^{-5}$~g/cm$^3$. The $\lambda_{tr}$ contribution fully determines the total thermal conductivity at the lowest temperatures considered here. Note that the electronic contribution can be neglected there because the ionization degree is virtually zero; see Fig.~\ref{benchmark_den}. The first peak in $\lambda$ at about 4000~K emerges due to the dissociation reaction heat conductivity of $\textrm{H}_2$ molecules in the PIP. This contribution becomes smaller at higher temperatures because most of the $\textrm{H}_2$ molecules are dissociated into H atoms. As temperature increases further, the H atoms are ionized, which leads to a second peak in the thermal conductivity at about 20~000~K due to the corresponding ionization reaction heat. A shoulder in $\lambda_r$ emerges at about 30~000~K due the ionization of He. The free electron density $n_e$ is systematically increasing with temperature so that $\lambda_e$ dominates the thermal conductivity $\lambda$ in the high-temperature limit above 25~000~K and both $\lambda_{tr}$ and $\lambda_r$ can be neglected there.

%%%%%%%%%%%%%%%%%%%%%%%%%%%%%%%%%%%%%%%%%%%%%%%%%%%%%%%
\section{Application to the atmosphere of the hot Jupiter HD~209458\MakeLowercase{b}}
\label{jupiter} 

%%% astrophysical introduction %%%
HD~209458b was the first exoplanet observed transiting its host star~\cite{Charbonneau_2000}.
With an orbital period of $3.5$~days, a semi-major axis of only 0.047~AU, a radius of $1.36~R_J$, 
and a mass of $0.69~M_J$, HD~209458b is clearly an inflated hot Jupiter~\cite{Sing2016}. 
Here $R_J$ and $M_J$ denote Jupiter's radius and mass, respectively. 

In this section we apply the methods discussed above to HD~209458b and discuss how the updated electrical conductivity 
would affect Ohmic heating. The electrical currents responsible for the Ohmic heating could penetrate down to a pressure level of few kbar according to B\&S10. We therefore focus the application of our PIP model on this pressure range and start with discussing the corresponding $P$-$T$ profile.

%% atmospheric modeling
\subsection{$P$-$T$ profile of the atmosphere}

\begin{figure}[!hbt] 
\centering  
\includegraphics[scale=0.6]{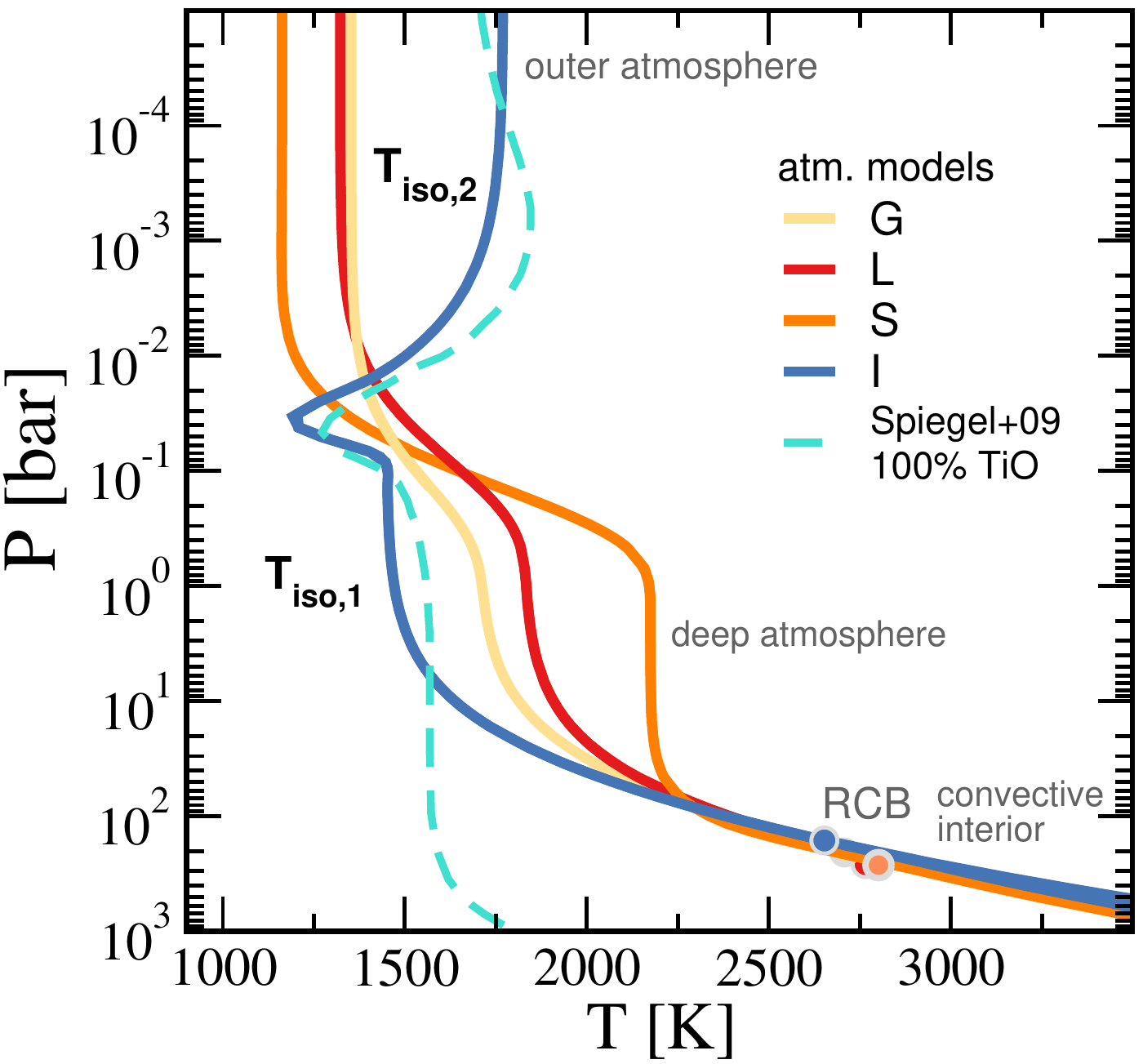}
\caption{Pressure-temperature profiles of the atmosphere of HD~209458b. Shown are the four atmospheric models used in this work, three without an inversion, namely G (yellow), L (red), and S (orange), and one with an inversion in the temperature, I (blue), located at about $0.03$~bar. Further features of the models are displayed: the location of the radiative-convective boundary (RCB), the onset of the convective interior, and the characteristic temperatures $\text{T}_{\text{iso},1}$ and $\text{T}_{\text{iso},2}$.}
\label{P_T_profile}	                       
\end{figure}

We calculate the composition and the transport coefficients of the planetary 
PIP for the four planetary models shown in Fig.~\ref{P_T_profile}.
The atmospheric models are obtained by fitting semi-analytical 1-d parametrizations 
to temperature-pressure ($P$-$T$) profiles suggested in the literature, 
following the approach by Poser \textit{et al.}~\cite{Poser_2019}. The 
parametrization guarantees a consistent description and allows us to extend 
all models to the same pressure range and to connect them to an adiabatic 
interior. 

Model G is based on the `globally averaged' theoretical $P$-$T$ curve by 
Guillot~\cite{Guillot_2010}, while model L replicates the most recent result 
by Line \textit{et al.}~\cite{Line_2016}, which is based on high-resolution 
spectroscopy data of the Hubble Space Telescope and data from the Spitzer 
Space Telescope for the planet's dayside. 
Both profiles turn out to be very similar. Profiles S and I follow 
suggestions by Spiegel~\textit{et al.}~\cite{Spiegel_2009}. While profile S 
has a particularly high temperature between $0.3$ and $100$~bar, model I, 
based on the variant with a solar abundance of TiO by \citet{Spiegel_2009}, 
shows a temperature inversion at pressures smaller than $30$~mbar. The reason 
is that the highly abundant TiO serves as an additional absorber in the upper 
atmosphere and leads to the rise in temperature.

%%%%%%%%%%%%%%%%%%%%%%%%%%%%%%%%%%%%%%%%%%%%%%%%%%%%%%%%%%%%%%%%%%%%%%%%%%%
%%%%%%%%%%%%%%%%%%%%%%%%%%%%%%%%%%%%%%%%%%%%%%%%%%%%%%%%%%%%%%%%%%%%%%%%%%%

%original vs. our parametrization of model I
Our parametrization of model I is broadly similar to the original profile of 
Spiegel~\textit{et al.}~\cite{Spiegel_2009} but assumes a shallower transition 
to the convective interior and, thus, predicts higher temperatures for pressures 
beyond $10$~bar. In addition, our temperatures are up to $100$~K lower than the 
original in the isothermal region between $1$ and $10$~mbar. Between $10^{-2}$ 
and $10^{-3}$~bar, the original shows a local maximum that is not present in our model. 
The temperatures in profile I are, therefore, up to $200\,$K colder than in the original paper. 

We connect our atmosphere profiles to an adiabatic interior model at the pressure level where the atmospheric temperature gradient matches with the adiabatic gradient. The respective transition points are marked with circles in Fig.~\ref{P_T_profile}. 
The interior model is derived from the usual structure equations for non-rotating, 
spherical gas planets; see e.g.~\citep{Nettelmann_2011}. Like B\&S10, we use a solar 
helium mass fraction of $Y=0.24$, assume no planetary core, and set the heavy-element 
mass fraction of both the atmosphere and the interior to the solar reference metallicity of $Z_\odot=0.015$~\cite{Lodders_2003}. For H and He we use the EOS of Saumon \textit{et al.}~\cite{SaumonChabrier1995}. Heavy elements are represented by the ice EOS of Hubbard \textit{et al.}~\cite{Hubbard_1989}. The upper boundary of our interior model is set to $P_{\rm{out}}(R_P)=10^{-2}$~bar. The heat flux from below is determined by the interior model (no core). The observed radius inflation is then obtained by adding extra energy during the thermal evolution~\cite{Poser_2019, ThorngrenFortney2018}.

B\&S10 also used variants of the original model I by \citet{Spiegel_2009} for their Ohmic 
dissipation study. Like us, they assumed a transition to an adiabatic interior model in a 
comparable pressure range. Their exact profiles have not been 
published but are likely similar to our model I. 

Beyond the radiative-convective boundary (RCB), the atmosphere models span a large temperature 
range of up to $750$~K around $1$~bar. This may partly be owed to the large local variation in 
brightness temperature with a dayside-to-nightside difference of about $500$~K~\citep{Zellem_2014} 
but mostly reflects the different model assumptions and a lack of observational 
constraints~\cite{Drummond+2020}. Note, however, that the most recent observation-based model 
by Line~\textit{et al.} \cite{Line_2016} could not confirm the inversion discussed by 
\citet{Spiegel_2009} and covers an intermediate temperature range. 

All of our atmosphere models have two nearly isothermal regions. The deeper region, labeled 
$\text{T}_{\text{iso},1}$ in Fig.~\ref{P_T_profile}, is a typical feature in strongly irradiated 
planets~\cite{Fortney+2008,Komacek_2017}. The shallower isothermal region $\text{T}_{\text{iso},2}$ 
from the $10$~mbar level to the outer boundary of our models is typical for analytical, 
semi-gray atmosphere models; see e.g.~\cite{Parmentier+2014}. For profiles G, L, and S, 
both regions are connected by a pronounced temperature drop of some hundred Kelvin. In the 
inversion profile, the temperature first drops but then again increases towards the outer boundary. 

%%% analysis of figures %%%
\subsection{Transport properties of the atmosphere}

\begin{figure}[!hbt] 
\centering   
\includegraphics[scale=0.6]{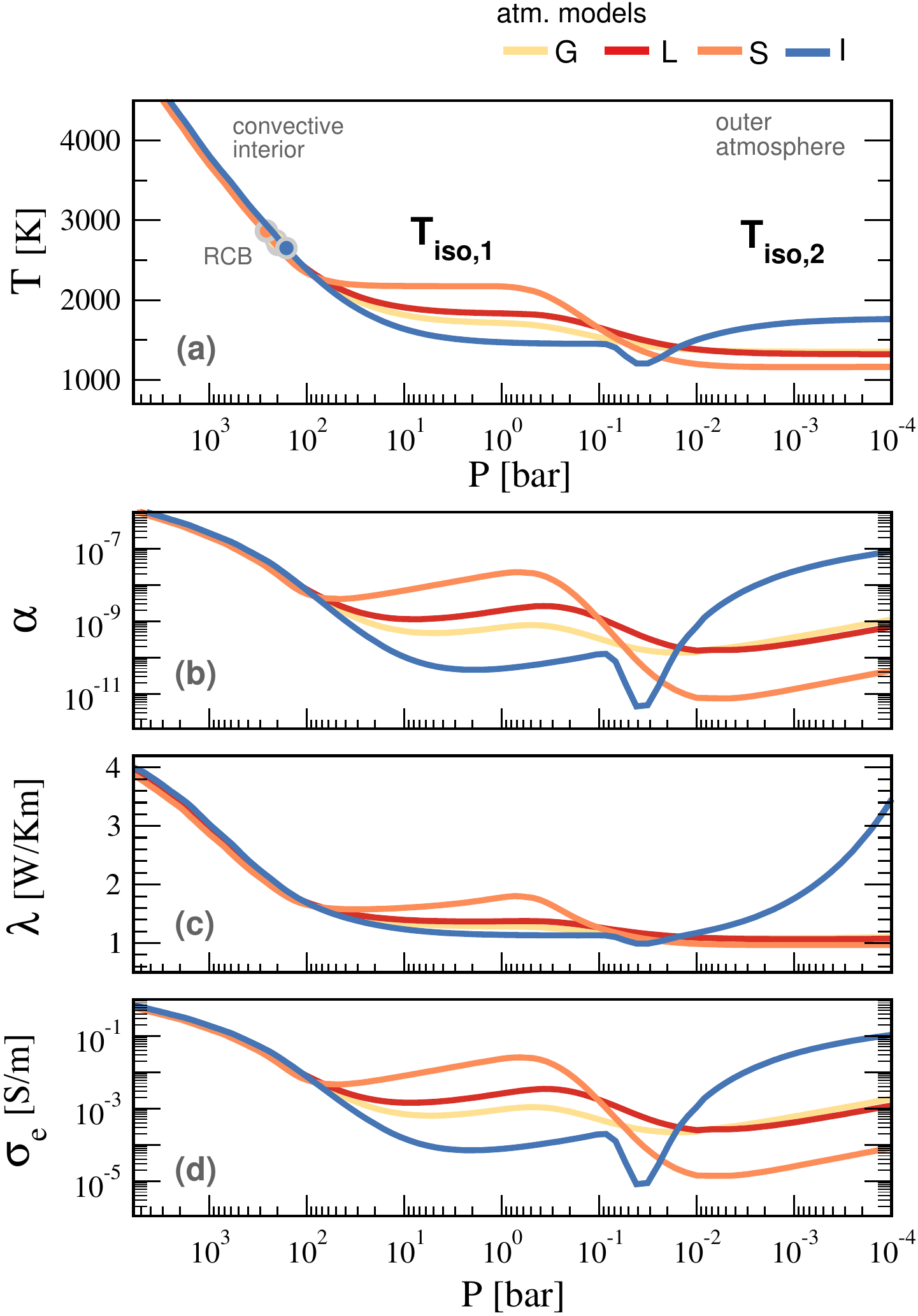}
\caption{Temperature $T$, ionization degree $\alpha$, thermal conductivity $\lambda$, and electrical conductivity $\sigma_e$ for different planetary interior models along the pressure axis of HD~209458b, specifically, for the four atmospheric models used in this work: G (yellow), L (red), and S (orange) as well as one with an inversion, I (blue). Circles in the temperature profile represent the location of the radiative-convective boundary (RCB).}
\label{HD209458b_TC}	                           
\end{figure}

We have calculated the ionization degree $\alpha$, the electrical conductivity 
$\sigma_e$, and the thermal conductivity $\lambda$ along our four $P$-$T$ models 
for HD~209458b; see Fig.~\ref{HD209458b_TC}. The ionization degree (panel b) and 
the electrical conductivity (panel d) are closely related and follow a very similar 
behavior; see Sec.~\ref{result_transport}. The thermal conductivity profile 
(panel c) also shows a similar form but with much smaller variations. 

%%%%%%%%%%%%%%%%%%%%%%%%%%%%%%%%%%%%%%%%%%%%%%%%%%%%%%%%%%%%%%%%%%%%%%%%%%%%%%%%
%%%%%%%%%%%%%%%%%%%%%%%%%%%%%%%%%%%%%%%%%%%%%%%%%%%%%%%%%%%%%%%%%%%%%%%%%%%%%%%%

In the two isothermal regions of each profile, the decreasing density causes $\alpha$ 
and $\sigma_e$ to increase outwards. However, the drastic changes of temperature 
in the intermediate regions between the isothermal layers influence $\alpha$ and 
$\sigma_e$ in more characteristic ways. This is especially the case in the inversion 
region in profile I (blue) where we find pronounced minima of $\alpha$ and $\sigma_e$ 
near $30\,$mbar; see Fig.~\ref{HD209458b_TC}. 

Due to the large differences between the models, the ionization degree and electrical 
conductivity differ by up to three orders of magnitude for the same pressure. The 
drop in electrical conductivity between the two isothermal regions varies from one 
order of magnitude in model G to more than three orders of magnitude in model S. 
The increase from the inner isothermal region to the RCB varies from a bit more
than two orders of magnitude in model S to four orders of magnitude in model I. 
In contrast, the variation of the thermal conductivity between the models is much 
smaller. The reason is that thermal conductivity is determined mostly by collisions 
between neutral particles in the relevant temperature range and is, thus, not 
susceptible to the strongly changing ionization degree.

\begin{figure}[!hbt] 
\centering  
\includegraphics[scale=0.6]{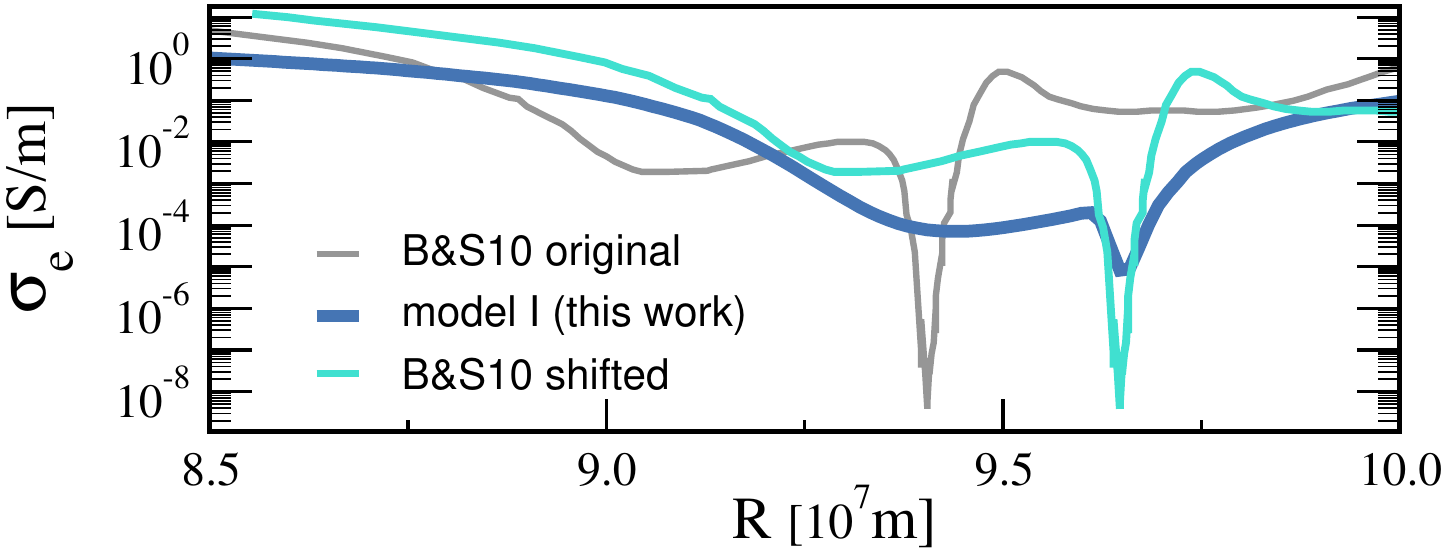}
\caption{Electrical conductivity $\sigma_e$ for model I along the radius axis of HD~209458b, compared to the results of Batygin and Stevenson ($B\&S10$)~\cite{Batygin_2010}. We show their original data (gray) as well as a   shifted version (cyan) to ease the comparison; see text. }
\label{BS_electricalcond_directcomparison}	                       
\end{figure}

Fig.~\ref{BS_electricalcond_directcomparison} compares the electrical conductivity for 
model I with the results taken from Fig.~2 in B\&S10. The associated pressure profile 
$P(r)$ is obtained by solving the equation of hydrostatic equilibrium. Each of the curves 
have a pronounced minimum in the electrical conductivity. Note that these minima are located 
at different radii in Fig.~\ref{BS_electricalcond_directcomparison}, which is likely caused 
by a different planetary radius assumed by B\&S10 that was, unfortunately, not stated in 
their paper. For better comparison, we also show a shifted B\&S10 profile in 
Fig.~\ref{BS_electricalcond_directcomparison} that aligns both minima. 

The electrical conductivity minimum predicted by B\&S10 is extraordinarily deep, with 
$\sigma_e$ dropping by six orders of magnitude. In contrast, the temperature dependence of our model (see Fig.~\ref{EC_vary_T}) yields a 
conductivity drop by only two orders of magnitude at the $300$~K temperature dip of our model I.  

In the inner isothermal region, the electrical conductivity is two orders of magnitude 
lower than suggested by B\&S10. Unfortunately, we do not know the exact atmosphere model 
used by B\&S10 but, as discussed above, it seems conceivable that the temperatures in this 
region are about $100$~K lower than assumed by B\&S10 for their Fig.~2. According to 
Fig.~\ref{EC_vary_T}, however, this would only explain a conductivity difference by about 
a factor $5$. At very low pressures and also toward the RCB, the electrical conductivities 
become more similar, likely because our model assumes higher temperatures. At the RCB, 
our electrical conductivity is about one order of magnitude lower than suggested by B\&S10.

%%%%%%%%%%%%%%%%%%%%%%%%%%%%%%
\subsection{Ohmic dissipation}

The electric currents $\vec{j}_e$ in the outer atmosphere are induced by 
the interaction of the fierce atmospheric winds with the planetary magnetic 
field according to Ohm's law:
\begin{equation}
\label{eq:Ohm}
  \vec{j}_e = \sigma_e\,\left(\vec{U}\times\vec{B}_0 - \nabla\Phi \right) \;,
\end{equation}
where $\vec{U}$ is the wind velocity, $\vec{B}_0$ the internally produced 
background field, and $\Phi$ the electric potential. Note that we use a fluid approach where the velocity describes the motion of the neutral medium (neutrals, ions, and electrons). Furthermore, we use a linear 
approximation, assuming that the magnetic field locally produced by the 
currents is smaller than the background field~\cite{Liu_2008, Wicht_2019a}. 
Using the fact that the currents are divergence-free, i.e., $\nabla\cdot\vec{j}_e=0$, 
allows calculating the missing electric potential (B\&S10). The global 
heating power from Ohmic dissipation is then simply given by the following
volume integral:
\begin{equation}
\label{eq:Heat}
 \dot{Q} = \int \frac{\vec{j}_e^2}{\sigma_e} dV \;.
\end{equation}

Being driven by the differential irradiation, the depth of the winds is limited~\citep{Perna_2012}. B\&S10 assume that they penetrate down to the $10$~bar level. Because the minimum in the electrical conductivity around $30$~mbar 
provides a boundary for the electric currents, only the layer from $10$~bar up to 
this minimum has to be considered for inducing the currents that could potentially penetrate deeper into the planet. We refer to this region as the \textit{induction layer}. 
While the electric currents in the induction layer already provide very powerful heating, the deeper penetrating currents are more relevant for explaining the inflation. We refer to the deeper layer where these currents remain significant as the \textit{leakage layer}, which may extend from $10$~bar to a few kbar (B\&S10). 

With no appreciable flows being present between $10$~bar and the RCB, the respective currents in the leakage layer obey the simpler relation:
\begin{equation}
\label{eq:Ohm2}
  \vec{j}_e = - \sigma_e\nabla\Phi\;.
\end{equation}
The electric potential differences $\nabla\Phi$ are determined by the 
action in the induction layer and the electrical conductivity distribution. B\&S10 therefore call the leakage layer the \textit{inert layer}. 
The electrical conductivity profile controls how deep the currents produced 
in the induction layer flow into the leakage layer. 

We can now roughly quantify the changes in Ohmic heating compared to B\&S10 
by simply rescaling their results with our new electrical conductivity profiles. 
B\&S10 assume a simple flow structure with typical velocities of $U=1$~km/s 
and a background field strength of $B_0=10$~Gauss. Because our electrical 
conductivity is about two orders of magnitude lower in the induction layer, 
the induced electric currents are two orders of magnitude weaker, according 
to Eq.~(\ref{eq:Ohm}). Consequently, the Ohmic heating power (\ref{eq:Heat}) 
is also two orders of magnitude lower. 

In the leakage layer, the currents encounter a conductivity that is more 
similar to the one assumed by B\&S10. Assuming that the conductivity is one 
order of magnitude lower (see Fig.~\ref{BS_electricalcond_directcomparison}),  
the deeper Ohmic heating is about $10^{-3}$ times smaller than in B\&S10. 
Explaining the inflation of HD~209458b requires a power of about $4\times10^{18}$~W 
to be deposited at or below the RCB~\citep{Burrows_2007}. While the models considered by B\&S10 deposit up to $10^{20}$~W in the convective 
interior, the lower the electrical conductivity of model I would render Ohmic 
heating too inefficient. 

However, as shown above, the Ohmic heating processes depend strongly on the conductivity and thus on the atmosphere model. Because of the higher temperatures, the electrical conductivity in the induction layers of the most up-to-date model L is comparable to that assumed by B\&S10; consequently the induced currents also have a similar magnitude. If assuming once more a ten times lower conductivity in the leakage layer, the leakage layer heating will be ten times stronger than in B\&S10, which is more than enough to explain the inflation. For the model S, the heating will be even stronger because of the particularly high temperatures in the induction region. 

Because the electrical currents depend linearly on the wind velocity $U$ and 
the background field strength $B_0$, the heating power (\ref{eq:Heat}) scales 
quadratically with both of these quantities. Updating the value of $U=1$~km/s 
assumed by B\&S10 with a newer estimate of $U=2$~km/s \cite{Snellen_2010}, 
thus, increases Ohmic heating by a factor of four. On the contrary, an 
indirect reassessment of the magnetic field strength of HD~209458b suggests 
that it may as well be in the order of $1$~Gauss \cite{Kislyakova_2014} 
rather than the $10$~Gauss assumed by B\&S10. This would reduce the Ohmic 
heating power by a factor of 100 and may render the process, once more, 
too inefficient to explain the inflation. 

All the estimates discussed above represent a linear approximation, 
assuming that the magnetic field produced by the locally induced currents 
is smaller than the background field in Eq.~(\ref{eq:Ohm}) 
\cite{Liu_2008,Wicht_2019a,Wicht_2019b}. The ratio of the locally induced 
field to the background field is roughly given by the magnetic Reynolds number 
\begin{equation}
 \mathrm{Rm}= U d \sigma_e \mu_0\;,
\end{equation}
where $\mu_0$ is the magnetic permeability of vacuum, 
and $d$ the electrical conductivity scale height:
\begin{equation}
 d = \frac{\sigma_e}{|\partial\sigma_e / \partial r|}\;. 
\end{equation} 
The linear approximation, therefore, breaks down when Rm exceeds one. 
When assuming $U=1$~km/s and the value $d=3\times10^2$~km as suggested 
by Fig.~\ref{BS_electricalcond_directcomparison}, this happens where 
the electrical conductivity is larger than $\sigma_e=10^{-2}$~S/m in 
the induction region. Model S, where $T_{\mathrm{iso,1}}=2200$~K, is the 
only model for which the linear approximation is certainly questionable.  

Observations suggest that dayside and nightside temperatures of HD~209458b 
differ by roughly $500$~K~\citep{Zellem_2014}. The fact that this difference 
is smaller than expected is, like the pronounced hotspot shift~\cite{Zellem_2014}, 
likely the result of heat distribution by the fierce winds in the upper atmosphere. 
The temperature dependence proposed here predicts that the electrical conductivity 
in the nightside induction region is about $10^3$ times lower than on the dayside. 
We, thus, expect that dayside heating would dominate.

%%%%%%%%%%%%%%%%%%%%%%%%%%%%%%%%%%%%%%%%%%%%%%%%%%%%%%%%%%%%%%%%%%%%%%%%%%
%%%%%%%%%%%%%%%%%%%%%%%%%%%%%%%%%%%%%%%%%%%%%%%%%%%%%%%%%%%%%%%%%%%%%%%%%%        
\section{Conclusions}
\label{concl} 

We have presented a model for calculating the chemical composition and 
electrical and thermal conductivity of low-density multicomponent plasmas 
suitable for applications in hot Jupiter atmospheres. This model is based 
on mass action laws and cross sections for all binary particle interactions
and generalizes an earlier model for the thermoelectric properties of 
one-component plasmas~\cite{Kuhlbrodt_2005} to multicomponent plasmas.
We have shown that the results for the ionization degree and, in particular, 
for the electrical conductivity can differ by several orders of magnitude 
from simpler models applied to hot Jupiter~\citep{Batygin_2010, Huang_2012} 
or hot Neptune atmospheres~\cite{Pu_2017}. 

Note that the plasma becomes nonideal with increasing depth (i.e.\ density), 
so that interaction contributions have to be treated when evaluating MALs 
for deeper atmosphere regions. Furthermore, simple expressions for the cross 
sections as used here no longer apply and the different scattering processes 
have to be treated on T matrix level by calculating the corresponding scattering 
phase shifts; see 
e.g.~\cite{Redmer_1997, Redmer_1999, Kuhlbrodt_2000, Kuhlbrodt_2005, Adams2010}. 
It would also be interesting to study the influence of the magnetic field of the 
planet on the transport properties, in particular for the hot and dilute outer 
atmosphere (ionosphere). This is subject of future work.

The plasma is strongly coupled and degenerate in the deep interior of the planet, 
so that first-principles approaches have to be applied in order to calculate the 
corresponding equation of state data, the ionization degree, and the transport 
properties. For instance, extensive molecular dynamics simulations have been 
performed for the ions in dense H-He plasmas in combination with electronic 
structure calculations using density functional theory (DFT-MD method). The 
corresponding results provide a reliable databases to determine interior 
profiles for density, temperature, and pressure~\cite{Nettelmann2008}, and 
to simulate the dynamo process based on further material properties such as 
electrical and thermal conductivity~\cite{French2012, Becker2018} for 
Jupiter~\cite{Gastine2014} and Jupiter-like planets. The deep interior is, 
however, not important for the study of Ohmic dissipation in the outer 
atmosphere so that the current results persist.

%%%
We have, therefore, used our results to predict the thermal and electrical 
conductivity for four different models proposed for the atmosphere of the hot 
Jupiter HD~209458b. The new estimates suggest that the electrical conductivity is 
between one and two orders of magnitude lower than assumed by B\&S10~\cite{Batygin_2010} 
in their study of Ohmic heating. While B\&S10 conclude that this additional heat 
source could explain the observed inflation of HD~209458b, our updated 
conductivities reduce the effect by up to three orders of magnitude and would 
make Ohmic heating too inefficient. 

However, newer internal models~\cite{Line_2016} suggest significantly higher 
temperatures in the planet's atmosphere than assumed for these estimates. 
The resulting higher electrical conductivity would guarantee more than enough 
Ohmic heat to explain the inflation, even for our lower electrical conductivity 
values. The large uncertainties in the atmospheric temperature, but also in 
the planet's magnetic field strength~\cite{Kislyakova_2014} yet prevent us to 
give reliable estimates of Ohmic heating in HD~209458b. 

Our estimates for the electric currents and, thus, for the Ohmic heating power 
largely follow simple scaling arguments based on previous 
attempts~\cite{Batygin_2010,Wicht_2019b}. It would be interesting to run 
refined numerical models that solve for electrical currents using the 
updated conductivities proposed here. Because of the significant radial and 
dayside-to-nightside variation in temperature, the electrical conductivity 
will also have a 3d field structure, making 3d simulations essential. 
Repeating the simplified calculations by B\&S10 would be a first step.
However, full magneto-hydrodynamic simulations are required should the 
locally induced magnetic fields and associated Lorentz forces prove important.

%%%%%%%%%%%%%%%%%%%%%%%%%%%%%%%%%%%%%%%%%%%%%%%%%%%%%%%%%%%%%%%%%%%%%%%% 
%%%%%%%%%%%%%%%%%%%%%%%%%%%%%%%%%%%%%%%%%%%%%%%%%%%%%%%%%%%%%%%%%%%%%%%%      
\section*{acknowledgments}
We thank Jan Maik Wissing for providing altitude data of the thermosphere and Nadine Nettelmann, Ludwig Scheibe, Martin Preising, and Clemens Kellermann for helpful discussions. This work was supported by the Deutsche Forschungsgemeinschaft (DFG) within the Priority Program 
SPP~1992 ``The Diversity of Exoplanets" and the Research Unit FOR~2440 ``Matter under Planetary Interior Conditions''. 
%%%%%%%%%%%%%%%%%%%%%%%%%%%%%%%%%%%%%%%%%%%%%%%%%%%%%%%%%%%%%%%%%%%
%%%%%%%%%%%%%%%%%%%%%%%%%%%%%%%%%%%%%%%%%%%%%%%%%%%%%%%%%%%%%%%%%%%

\end{document}